\documentclass[aps,pre,reprint]{revtex4-2}

\usepackage{amsmath,amssymb,textcomp,graphicx}

\begin{document}

\title{Deformation of a magnetic liquid drop and unsteady flow inside and outside it in a non-stationary magnetic field}	
\author{Alexander~N.~Tyatyushkin} 
\affiliation{Institute of Mechanics, Moscow State University, Michurinskiy Ave., 1, Moscow 119192, Russia} 

\begin{abstract}%
Small steady-state deformational oscillations of a drop of a viscous magnetic liquid in a non-stationary uniform magnetic field are theoretically investigated for low Reynolds numbers. The drop is suspended in another viscous magnetic liquid immiscible with the former. The non-stationary magnetic field causes flow inside and outside the drop. The inertia is taken into account by regarding the flow as essentially unsteady. The variation of the magnetic field is so slow that the approximation of quasi-stationary magnetic field may be used. 
\end{abstract}

\newcommand{\vecHa}{\vec{H}_\mathrm{a}}
\newcommand{\Ha}{H_\mathrm{a}}
\newcommand{\rhoi}{\rho_\mathrm{i}}
\newcommand{\etai}{\eta_\mathrm{i}}
\newcommand{\mui}{\mu_\mathrm{i}}
\newcommand{\rhoe}{\rho_\mathrm{e}}
\newcommand{\etae}{\eta_\mathrm{e}}
\newcommand{\mue}{\mu_\mathrm{e}}
\newcommand{\Rd}{a}
\newcommand{\sigmas}{\sigma_\mathrm{s}}
\newcommand{\iu}{\mathrm{i}}%
\newcommand{\re}{\mathop{\mathrm{Re}}}%
\newcommand{\en}{\mathrm{e}}%
\newcommand{\kappai}{\varkappa_\mathrm{i}}
\newcommand{\kappae}{\varkappa_\mathrm{e}}

\maketitle
\section*{Introduction.}
A non-stationary magnetic field applied to a drop of a magnetic liquid suspended in an ordinary liquid or to a drop of an ordinary liquid suspended in a magnetic liquid causes a motion of the liquids inside and outside the drop. Investigation of this phenomenon is interesting from the point of view of application to the actuation of motion of liquids in various devices including microfluidic ones. Besides, the study of this phenomenon is interesting from the point of view of basic science.

Results of experimental and theoretical investigation of the behavior of magnetic liquid drops in non-stationary magnetic fields are presented in works \cite{Subkhankulov,Bacri--94,Lebedev--Morozov,Morozov,Bratukhin--,Morozov--Lebedev,Bratukhin--Lebedev,Lebedev--,Stikuts--,Langins--}. The behavior of an ordinary liquid drop suspended in a magnetic liquid was also studied \cite{Dikanskiy--Zakinyan}.

In works \cite{Bacri--94,Lebedev--Morozov,Morozov},  \cite{Morozov--Lebedev}, and \cite{Lebedev--} for the theoretical description of the behavior of a magnetic liquid drop, models based on rather strong assumptions about the shape of the drop and about the flow inside and outside it were used. In works \cite{Stikuts--} and \cite{Langins--} the tree-dimensional boundary element method algorithm was used for the numerical calculations. In works \cite{Subkhankulov}, \cite{Bratukhin--}, \cite{Bratukhin--Lebedev}, and \cite{Stikuts--} asymptotic methods were used. In  the present work the problem about the shape of the drop and flow inside and outside it is also solved with asymptotic methods without the use of any assumptions about them. Within this approach, the shape of the drop is determined by equations obtained with the help of asymptotic methods from the system of equations and boundary conditions that determine the magnetic field and the flow. In order to provide the applicability of the asymptotic methods, the scope of the present theoretical investigation is restricted by the case of weakly deformed drops. The unsteady flow inside and outside a drop of a viscous magnetic liquid suspended in another viscous magnetic liquid in non-stationary magnetic field is investigated with taking into account the influence of form of the drop on the field and flow. This approach was used also in works \cite{EPJ,MHD18,BRAS-Ph}, in which the behavior of a magnetic liquid drop in non-stationary fields for low Reynolds numbers was investigated within the quasi-steady flow approximation, and in \cite{FD21}, in which the Reynolds numbers were regarded as high and the viscosity was neglected. Thus, the inertia was completely neglected in \cite{EPJ,MHD18,BRAS-Ph}. In the present work, the inertia is taken into account simultaneously with the viscosity by refusing the quasi-steady flow approximation.

The Reynolds number in the problem about deformational oscillations of a drop under action of non-stationary fields tends to zero as the amplitude of the oscillation tends to zero. Indeed, the velocity in that case has the order of magnitude $\omega\Delta{\Rd}$, where $\omega$ is the angular frequency of the oscillations of the applied field, $\Delta{\Rd}$ is the amplitude of the deformational oscillations of the drop (maximal variation of the distance from the center of the drop to its surface). Then the Reynolds number tends to zero as the amplitude of the oscillations tends to zero. Thus, in the case of small oscillations of the drop, the influence of the convection term in the momentum equation can be neglected, and the influence of the inertia can be described by the unsteady therm. This is directly confirmed by the fact that, in the problem about deformational oscillations of a drop in the case of when the influence of the viscosity is completely neglected (see \cite{FD21}), the influence of the convection term in the correction of the first order over the parameter the smallness of which provides the smallness of the deformations is absent, although it is taken into account in the input system of equations. It is also followed from this fact that the solution of the problem about small oscillations of the drop with taking into account both the viscosity and inertia should tend to the solution in \cite{FD21} as the maximal viscosity of the liquids tends to zero with exception of the case when the frequency of the applied field is close to the resonance frequency.

In order to use the asymptotic methods for the investigation of small oscillations of drops under action of non-stationary fields developed in the works \cite{EPJ,MHD18,BRAS-Ph,FD21} with the use of unsteady term in the momentum equation, it is necessary to obtain the general solution to the system of hydrodynamic equations that contains this term. This solution was obtained and is written down below. The problem stated above presents a good opportunity to use this general solution since the problems solved in the works \cite{EPJ,MHD18,BRAS-Ph,FD21} allow checking the results for the limit cases when the influence of either the inertia or the viscosity is neglected.


\section{Setting of the problem.}

\subsection{Object of investigation.}

Consider a drop of incompressible magnetic liquid in an applied non-stationary uniform magnetic field with intensity $\vecHa=\vecHa(t)$, where $t$ is the time. The density, viscosity, and magnetic permeability of the liquid inside the drop are $\rhoi$, $\etai$, and $\mui$. The radius of the drop in the undeformed state is $\Rd$. The drop is suspended in an incompressible magnetic liquid with the density $\rhoe$, viscosity $\etae$, and magnetic permeability $\mue$. The surface tension of the interface between the liquids is $\sigmas$. The liquids are regarded as sufficiently viscous so that the low Reynolds number approximation is valid. The influence of the inertia of the liquids is taken into account by means of regarding the flow as essentially unsteady. The variation of the magnetic field is so slow and the conductivities of the liquids are so small that the ferrohydrodynamics approximation is valid (see \cite{Rosensweig}).

\subsection{System of equations.}

\newcommand{\tsigmae}{\hat\sigma_{\rm m}}%
\newcommand{\tsigmah}{\hat\sigma_{\rm v}}%
In order to find the magnetic field intensity, $\vec{H}$, velocity, $\vec{v}$, and pressure, $p$, as functions of the radius-vector, $\vec{r}$, and time, $t$, as well as the shape of the drop under the above made assumptions, the system of equations of ferrohydrodynamics written down in the quasi-stationary magnetic field and low Reynolds number approximations is used. This system consists of the continuity equation for an incompressible liquid
\begin{equation}
\nabla\cdot\vec v = 0
,
\label{ce}
\end{equation}
the motion equation in the low Reynolds number approximation for arbitrary Strouhal numbers
\begin{equation}
\rho\dfrac{\partial\vec{v}}{\partial{t}}
=
\nabla\cdot
\left(-p \hat I + \hat\sigma_\mathrm{m} + \hat\sigma_\mathrm{v}\right)
,
\label{me0}
\end{equation}
the Maxwell's equations in the ferrohydrodynamics and quasi-stationary field approximations
\begin{equation}
\nabla \cdot \vec B = 0
,\qquad
\nabla \times \vec H = 0
,
\label{Me}
\end{equation}
and the constitutive relation 
\begin{equation}
\vec B = \mu \vec H
.
\label{BH}
\end{equation}
Here, the formulas are written down for Gaussian system of units, $\cdot$ and $\times$ denote the scalar and vector products, $\vec{B}$ is the magnetic induction, $\rho=\rhoi$, $\eta=\etai$, and $\mu=\mui$ inside the drop, $\rho=\rhoe$, $\eta=\etae$, and  $\mu=\mue$ outside it, $\nabla$ is the nabla operator, $\hat{I}$ is the identity tensor, $\tsigmae$ and $\tsigmah$ are the tensors of magnetic and viscous stresses expressed as follows
\begin{eqnarray}
\tsigmae
&=&
\frac{1}{4\pi} \vec B \vec H
-
\frac{1}{8\pi} \vec B \cdot \vec H \hat I
,
\label{sigmae}
\\*
\tsigmah
&=&
2 \eta \left(\nabla \vec v\right)^\mathrm{S}
,
\label{sigmah}
\end{eqnarray}
where $\vec{A}\vec{B}$ denotes the dyadic product of the vectors $\vec{A}$ and $\vec{B}$, $\nabla\vec{f}$ denotes the dyadic product of the nabla operator and the vector field $\vec{f}=\vec{f}(\vec{r})$, $\hat{T}^\mathrm{S}$ denotes the symmetric part of the tensor $\hat{T}$.

Using the continuity equation Eq.~(\ref{ce}) and Maxwell's equations Eq.~(\ref{Me}) and taking into account that the magnetic permeability is uniform, the motion equation Eq.~(\ref{me0}) can be rewritten in the form of the Navier--Stokes equation in the low Reynolds number approximation for arbitrary Strouhal numbers
\begin{equation}
\rho\dfrac{\partial\vec{v}}{\partial{t}}
=
-
\nabla{p}
+
\eta\Delta\vec{v}
,
\label{me}
\end{equation}
where $\Delta$ is the Laplacian.

\subsection{Boundary conditions.} 

The boundary conditions on the interface between the liquids include the impenetrability condition 
\begin{equation}
\left.\vec v\right|_{\rm e}\cdot\vec{n}
=
\left.\vec v\right|_{\rm i}\cdot\vec{n}
=
{v}_{\mathrm{s}n}
,
\label{BC:vn}
\end{equation}
the no-slip condition
\begin{equation}
\left[\vec v\right]_{\rm s}\times\vec n=0
,
\label{BC:vt}
\end{equation}
the conditions for the jumps of the normal and tangential components of the stress vector $\vec\sigma_n=\vec{n}\cdot\left(-p\hat{I}+\tsigmae+\tsigmah\right)$
\begin{eqnarray}
\vec{n}\cdot
\left[
 -p \hat I + \tsigmae + \tsigmah
\right]_{\rm s}
\cdot\vec{n}
&=&
-
\sigma_{\rm s} \mathcal{H}
,
\label{BC:sn}
\\*
\vec{n}\cdot
\left[
 \tsigmae + \tsigmah
\right]_{\rm s}
\times\vec{n}
&=&
0
,
\label{BC:st}
\end{eqnarray}
the continuity conditions for the tangential component of the magnetic field intensity
\begin{equation}
\left[\vec H\right]_{\rm s}\times\vec{n}=0
\label{BC:H}
\end{equation}
and for the normal component of the magnetic field induction
\begin{equation}
\left[\vec{B}\right]_{\rm s}\cdot\vec{n}
=
0
.
\label{BC:B}
\end{equation}
Here, $\left.A\right|_{\textrm{i}}$ and $\left.A\right|_{\textrm{e}}$ denote the values of the quantity $A$ on the interface between the liquids approached from inside and outside the drop, respectively,
$[A]_{\rm s}=\left.A\right|_{\rm e}-\left.A\right|_{\rm i}$
denotes the jump of the quantity $A$ at the interface when moving from the inside to the outside, $\nabla_\mathrm{s}$ denotes the surface nabla operator, $\vec{n}$ is the external normal unit vector at a given point of the interface, ${v}_{\mathrm{s}n}$ is the normal component of the velocity of the surface of the drop at a given point, $\mathcal{H}$ is the mean curvature at a given point of the surface of the drop. Note that definition of the mean curvature used in the present work is such that it takes negative values on the surface of a convex domain.

The continuity conditions for the tangential components of the velocity and electric field intensity allow determining the following vector fields defined on the surface of the drop
\begin{equation}
\vec{v}_{\mathrm{s}\tau}
=
\vec{n}\times
\left(\left.\vec{v}\right|_\mathrm{i}\times\vec{n}\right)
=
\vec{n}\times
\left(\left.\vec{v}\right|_\mathrm{e}\times\vec{n}\right)
,
\label{d:vs}
\end{equation}
\begin{equation}
\vec{H}_\mathrm{s}
=
\vec{n}\times
\left(\left.\vec{H}\right|_\mathrm{i}\times\vec{n}\right)
=
\vec{n}\times
\left(\left.\vec{H}\right|_\mathrm{e}\times\vec{n}\right)
.
\label{d:Es}
\end{equation}

The boundary conditions at infinity have the form
\begin{align}
\vec v &\to 0 &\quad\text{as}\quad r &\to \infty
,
\label{BC:inf:v}
\\*
p &\to p_\infty
&\quad\text{as}\quad r &\to \infty
,
\label{BC:inf:p}
\\*
\vec H &\to \vecHa
&\quad\text{as}\quad r &\to \infty
,
\label{BC:inf:H}
\end{align}
where $p_\infty$ is the pressure at infinity.

Besides, $\vec{v}(\vec{r},t)$, $p(\vec{r},t)$, and $\vec{H}(\vec{r},t)$ should be bounded for all the bounded values of $\vec{r}$.

\section{Solution.}
\subsection{Deformation of the drop.} 

Let the surface of the drop be given by the following equation
\begin{equation}
r
=
\left|\vec{r}\right|
=
\Rd
+
h\left(\dfrac{\vec{r}}{{r}},t\right)
.
\label{eq:S}
\end{equation}
In order the deformations of the drop to be small, the following condition should be fulfilled
\begin{equation}
h
=
h\left(\dfrac{\vec{r}}{{r}},t\right)
\ll
\Rd
.
\end{equation}
The function $h\left(\vec{r}/{r},t\right)$ can be represented in the form 
\begin{equation}
h
=
h\left(\dfrac{\vec{r}}{{r}},t\right)
=
\sum\limits_{n=2}^\infty
 \hat{h}_n\stackrel{n}{\cdot}\dfrac{\vec{r}\,^n}{{r}^n}
.
\label{h}
\end{equation}
Here, $\hat{h}_n=\hat{h}_n(t)$ ($n>1$) are some tensors each of which is an arbitrary irreducible tensor of $n$th order, i.e., a tensor of $n$th order symmetric with respect to any pair of indices and such that its contraction with the identity tensor over any pair of indices is equal to zero (see \cite{LL2}). Here and in what follows, $\vec{b}\,^n$ denotes $n$th dyadic degree of the vector $\vec{b}$ (i.e., the $(n-1)$-multiple dyadic product of the vector $\vec{b}$ by itself), $\stackrel{n}{\cdot}$ denotes $n$-multiple contraction of tensors over the adjacent indices.

The impenetrability condition (\ref{BC:vn}) yields the following equation for the function $h\left(\vec{r}/{r},t\right)$
\begin{equation}
\vec{n}\cdot\dfrac{\vec{r}}{{r}}
\dfrac{\partial{h}}{\partial{t}}
=
{v}_{\mathrm{s}n}
.
\label{vsn-h}
\end{equation}

\subsection{Magnetic field.}

The intensity of the magnetic field is sought for in the form
\begin{equation}
\vec{H}
=
\nabla\psi
,
\label{H}
\end{equation}
where 
\begin{equation}
\psi
=
\begin{cases}
\vec{H}_\mathrm{a}\cdot\vec{r}
-
\displaystyle
\sum\limits_{n=1}^\infty
 \hat{M}_{\mathrm{e}n} \stackrel{n}{\cdot}
 \dfrac{\vec{r}\,^n}{r^{2n+1}}
,\ 
r>\Rd+h
,
\\
\vec{H}_\mathrm{a}\cdot\vec{r}
-
\displaystyle
\sum\limits_{n=1}^\infty
 \hat{M}_{\mathrm{i}n} \stackrel{n}{\cdot}
 \dfrac{\vec{r}\,^n}{\Rd^{2n+1}}
,\ 
r\leqslant\Rd+h
.
\end{cases}
\label{psi}
\end{equation}
Here, $\hat{M}_{\mathrm{e}n}=\hat{M}_{\mathrm{e}n}(t)$ and $\hat{M}_{\mathrm{i}n}=\hat{M}_{\mathrm{i}n}(t)$ are tensors which are either some vectors for $n=1$ or, for $n>1$, some irreducible tensors of $n$th order. For a field given in this form, Maxwell's equations Eq.~(\ref{Me}), the constitutive relation Eq.~(\ref{BH}), the condition at infinity Eq.~(\ref{BC:inf:H}), and the condition of boundedness of $\vec{H}$ are automatically satisfied.

\subsection{Flow.}

The velocity and pressure in the flow are sought for in the form
\begin{multline}
\vec{v}
=
\vec{v}(\vec{r},t)
=
\dfrac{1}{2\pi}
\int\limits_{-\infty}^\infty
 \en^{-\mathrm{i}\omega{t}}\vec{v}_{1\omega}
\mathrm{d}\omega
\\
+
\dfrac{1}{2\pi}
\int\limits_{-\infty}^\infty
 \en^{-\mathrm{i}\omega{t}}\vec{v}_{2\omega}
\mathrm{d}\omega
,
\label{v:gs}
\end{multline}
\begin{multline}
{p}
=
{p}(\vec{r},t)
=
\dfrac{1}{2\pi}
\int\limits_{-\infty}^\infty
 \en^{-\mathrm{i}\omega{t}}{p}_{\omega}
\mathrm{d}\omega
\\
+
\begin{cases}
{p}_\infty
,
\quad
r>\Rd+h
,
\\
{p}_0
,\quad
r<\Rd+h
,
\end{cases}
\end{multline}
\begin{equation}
\vec{v}_{1\omega}
=
\begin{cases}
\vec{v}_{1\mathrm{e}\omega}
,\quad 
r>\Rd+h
,
\\
\vec{v}_{1\mathrm{i}\omega}
,\quad 
r\leqslant\Rd+h
,
\end{cases}
\end{equation}
\begin{multline}
\vec{v}_{1\mathrm{e}\omega}
=
\nabla\times
 \left(
  \left\{
   \sum_{n=1}^\infty
   \left[
    \dfrac
     {{Q}_{-n-1}\left(\kappae\dfrac{r}{\Rd}\right)}
     {\kappae^2}
    \hat{B}_{\mathrm{e}n}
    +
    \hat{C}_{\mathrm{e}n}
   \right]
 \rule{0em}{5ex}
 \right.
\right.
\\
\left.
 \left.
 \rule{0em}{5ex}
   \stackrel{n-1}{\cdot}
   \dfrac{\vec{r}\,^{n-1}{a}^{n+2}}{{r}^{2n+1}}
  \right\} 
  \times\vec{r}
 \right)
,
\end{multline}
\begin{multline}
\vec{v}_{1\mathrm{i}\omega}
=
\nabla\times
 \left(
  \left\{
   \sum_{n=1}^\infty
   \left[
    \dfrac
     {{Q}_{n}\left(\kappai\dfrac{r}{\Rd}\right)}
     {\kappai^2}
    \hat{B}_{\mathrm{i}n}
    +
    \hat{C}_{\mathrm{i}n}
   \right]
 \rule{0em}{5ex}
 \right.
\right.
\\
\left.
 \left.
 \rule{0em}{5ex}
   \stackrel{n-1}{\cdot}\dfrac{\vec{r}\,^{n-1}}{{a}^{n-1}}
  \right\}
  \times\vec{r}
 \right)
,
\end{multline}
\begin{equation}
\vec{v}_{2\omega}
=
\begin{cases}
\vec{v}_{2\mathrm{e}\omega}
,
\quad
r>\Rd+h
,
\\
\vec{v}_{2\mathrm{i}\omega}
,\quad
r\leqslant\Rd+h
,
\end{cases}
\end{equation}
\begin{multline}
\vec{v}_{2\mathrm{e}\omega}
=
\sum_{n=1}^\infty
n
\left[
 \kappae^{2n+1}
 {G}_{-n-1}\left(\kappae\dfrac{r}{\Rd}\right)
 \hat\Omega_{\mathrm{e}n}
 \rule{0em}{3ex}
\right.
\\
\left.
 \rule{0em}{3ex}
 \stackrel{n-1}{\cdot}\dfrac{\vec{r}\,^{n-1}}{{a}^{n-1}}
\right]
\times\vec{r}
,
\end{multline}
\begin{equation}
\vec{v}_{2\mathrm{i}\omega}
=
\sum_{n=1}^\infty
n
\left[
 {G}_{n}\left(\kappai\dfrac{r}{\Rd}\right)
 \hat\Omega_{\mathrm{i}n}
 \stackrel{n-1}{\cdot}\dfrac{\vec{r}\,^{n-1}}{{a}^{n-1}}
\right]
\times\vec{r}
,
\end{equation}
\begin{equation}
{p}_{\omega}
=
\begin{cases}
{p}_{\mathrm{e}\omega}
,
\quad
r>\Rd+h
,
\\
{p}_{\mathrm{i}\omega}
,
\quad
r<\Rd+h
,
\end{cases}
\end{equation}
\begin{equation}
{p}_{\mathrm{e}\omega}
=
\dfrac{\etae}{\Rd}
\left[
 2(2n-1)\hat{B}_{\mathrm{e}n}
 +
 \kappae^2\hat{C}_{\mathrm{e}n}
\right]
\stackrel{n}{\cdot}\dfrac{\vec{r}\,^{n}{a}^{n+1}}{{r}^{2n+1}}
,
\end{equation}
\begin{equation}
{p}_{\mathrm{i}\omega}
=
\dfrac{\etai}{\Rd}
\left[
 \dfrac{2(n+1)(2n+3)}{n}
 \hat{B}_{\mathrm{i}n}
 -
 \kappai^2\dfrac{(n+1)}{n}\hat{C}_{\mathrm{i}n}
\right]
\stackrel{n}{\cdot}\dfrac{\vec{r}\,^{n}}{{a}^{n}}
,
\end{equation}
\begin{equation}
\kappae
=
(1-\mathrm{i})\sqrt{\dfrac{\rhoe\omega\Rd^2}{2\etae}}
,\quad
\kappai
=
(1-\mathrm{i})\sqrt{\dfrac{\rhoi\omega\Rd^2}{2\etai}}
,
\end{equation}
\begin{equation}
{G}_{n+1}(z)
=
\dfrac{2n+3}{z}{G}'_{n}(z)
,\quad
{G}_{0}(z)
=
\dfrac{\sinh{z}}{z}
,
\end{equation}
\begin{equation}
{G}_{-n-2}(z)
=
-
\dfrac{1}{(2n+1)z}{G}'_{-n-1}(z)
,\quad
{G}_{-1}(z)
=
\dfrac{\en^{-{z}}}{z}
,
\end{equation}
\begin{equation}
{Q}_{n}\left(z\right)
=
2(2n+3)
\left[
 {G}_{n}\left(z\right)
 -
 1
\right]
,
\quad
n\geqslant0
,
\end{equation}
\begin{equation}
{Q}_{-n-1}\left(z\right)
=
-
2(2n-1)
\left[
 z^{2n+1}
 {G}_{-n-1}\left(z\right)
 -
 1
\right]
,
\quad
n\geqslant0
.
\label{Q-n-1:gs}
\end{equation}
Here, $\hat\Omega_{\mathrm{e}n}=\hat\Omega_{\mathrm{e}n,\omega}$, $\hat{B}_{\mathrm{e}n}=\hat{B}_{\mathrm{e}n,\omega}$, $\hat{C}_{\mathrm{e}n}=\hat{C}_{\mathrm{e}n,\omega}$, $\hat\Omega_{\mathrm{i}n}=\hat\Omega_{\mathrm{i}n,\omega}$, $\hat{B}_{\mathrm{i}n}=\hat{B}_{\mathrm{i}n,\omega}$, $\hat{C}_{\mathrm{i}n}=\hat{C}_{\mathrm{i}n,\omega}$ are some tensors depending on $\omega$ each of which is either an arbitrary vector for $n=1$ or an arbitrary irreducible tensor for $n>1$, $p_0=p_0(t)$ is the pressure at the center of the drop, which is to be found. For a flow given in this form, the continuity equation Eq.~(\ref{ce}), the motion equation Eq.~(\ref{me}), the conditions at infinity Eqs.~(\ref{BC:inf:v})--(\ref{BC:inf:p}), and the condition of boundedness of $\vec{v}$ and ${p}$ are automatically satisfied.

The general solution Eqs.~(\ref{v:gs})--(\ref{Q-n-1:gs}) of the equations Eq.~(\ref{ce})--(\ref{me}) was obtained as the generalization of the general Lamb's solution for the case of unsteady flows with arbitrary Strouhal numbers. It turns into the general Lamb's solution (cf. \cite{Lamb}, art. 336 and \cite{EPJ}) for the quasi-steady flow approximation. 

\subsection{Asymptotic expansion.}

Thus, in order to solve the problem set above, it is necessary to find the unknown vector and tensor functions $\hat{M}_{\mathrm{e}n}=\hat{M}_{\mathrm{e}n}(t)$, $\hat\Omega_{\mathrm{e}n}=\hat\Omega_{\mathrm{e}n,\omega}$, $\hat{B}_{\mathrm{e}n}=\hat{B}_{\mathrm{e}n,\omega}$, $\hat{C}_{\mathrm{e}n}=\hat{C}_{\mathrm{e}n,\omega}$, $\hat{M}_{\mathrm{i}n}=\hat{M}_{\mathrm{i}n}(t)$, $\hat\Omega_{\mathrm{i}n}=\hat\Omega_{\mathrm{i}n,\omega}$, $\hat{B}_{\mathrm{i}n}=\hat{B}_{\mathrm{i}n,\omega}$, and $\hat{C}_{\mathrm{i}n}=\hat{C}_{\mathrm{i}n,\omega}$  ($n=1,2,\dots$) as well as the tensor functions $\hat{h}_n=\hat{h}_n(t)$ ($n=2,\dots$) and the scalar function $p_0=p_0(t)$ using the remaining boundary conditions Eqs.~(\ref{BC:vn})--(\ref{BC:B}), and the equation Eq.~(\ref{vsn-h}). The functions are sought for in the form of the following asymptotic expansions over the parameter 
\begin{equation}
\alpha
=
\dfrac{9\Rd\mue{H}_\mathrm{am}^2}{32\pi\sigmas}
, 
\end{equation}
\begin{multline}
\hat{M}_{\mathrm{e}n}
\sim
\sum_{j=0}^\infty \alpha^j \hat{M}_{\mathrm{e}n,j}
,\ 
\hat\Omega_{\mathrm{e}n}
\sim
\sum_{j=1}^\infty \alpha^j \hat\Omega_{\mathrm{e}n,j}
,\ 
\\*
\hat{B}_{\mathrm{e}n}
\sim
\sum_{j=1}^\infty \alpha^j \hat{B}_{\mathrm{e}n,j}
,\ 
\hat{C}_{\mathrm{e}n}
\sim
\sum_{j=1}^\infty \alpha^j \hat{C}_{\mathrm{e}n,j}
,\ 
\\*
\hat{M}_{\mathrm{i}n}
\sim
\sum_{j=0}^\infty \alpha^j \hat{M}_{\mathrm{i}n,j}
,\ 
\hat\Omega_{\mathrm{i}n}
\sim
\sum_{j=1}^\infty \alpha^j \hat\Omega_{\mathrm{i}n,j}
,\ 
\\*
\hat{B}_{\mathrm{i}n}
\sim
\sum_{j=1}^\infty \alpha^j \hat{B}_{\mathrm{i}n,j}
,\ 
\hat{C}_{\mathrm{i}n}
\sim
\sum_{j=1}^\infty \alpha^j \hat{C}_{\mathrm{i}n,j}
,\ 
\\*
{p}_0
\sim
\sum_{j=0}^\infty \alpha^j {p}_{0j}
,\ 
\hat{h}_n
\sim
\sum_{j=1}^\infty \alpha^j \hat{h}_{n,j}
\text{ as }
\alpha\to0
.
\label{asym}
\end{multline}
Here, ${H}_\mathrm{am}$ is the maximal absolute value of the intensity vector of the applied magnetic field. The parameter $\alpha$ is the magnetic Bond number for the present problem.

\subsection{Steady-state oscillations.}

For the forced steady-state oscillations, the found nonzero terms of the asymptotic expansions are as follows
\begin{equation}
\hat{M}_{\mathrm{e}1,0}
=
\vec{M}_{\mathrm{e}1,0}
=
\hat{M}_{\mathrm{i}1,0}
=
\vec{M}_{\mathrm{i}1,0}
=
\dfrac{\mui-\mue}{\mui+2\mue}\Rd^3\vec{H}_\mathrm{a}
,
\end{equation}
\begin{equation}
p_{0,0}
=
p_\infty
+
\dfrac{2\sigmas}{\Rd}
+
\dfrac{3\mue}{8\pi}
\dfrac{\mue-\mui}{\mui+2\mue} 
\Ha^2
,
\end{equation}
\begin{multline}
\hat{B}_{\mathrm{e}2,1}
=
-
\dfrac{\mathrm{i}\omega\tau}{2}\hat{h}_{2,1,\omega}
\left[
   14\dfrac{\etai}{\etae}
   \dfrac
    {{Q}_{2}'\left(\kappai\right)}
    {\kappai}
 +
 5\dfrac{\etai}{\etae}
 {Q}_{2}''\left(\kappai\right)
\right.
\\
\shoveright{
\left.
 +
 16
 \dfrac
  {{Q}_{2}'\left(\kappai\right)}
  {\kappai}
\right]
}
\\
\left(
  3
  \left\{
   \dfrac{\etai}{\etae}
   \dfrac
    {{Q}_{-3}'\left(\kappae\right)}
    {\kappae}
   \left[
    4\dfrac{{Q}_{2}'\left(\kappai\right)}{\kappai}
    +
    {Q}_{2}''\left(\kappai\right)
   \right]
 \right.
\right.
\\
\left.
 \left.
   +
   \dfrac
    {{Q}_{2}'\left(\kappai\right)}
    {\kappai}
   \left[
    6\dfrac{{Q}_{-3}'\left(\kappae\right)}{\kappae}
    -
    {Q}_{-3}''\left(\kappae\right)
   \right]
  \right\}
\right)^{-1}
,
\label{Be21}
\end{multline}
\begin{multline}
\hat{C}_{\mathrm{e}2,1}
=
\dfrac{\mathrm{i}\omega\tau}{2}\hat{h}_{2,1,\omega}
\left\{
   \dfrac{\etai}{\etae}
   \dfrac
    {{Q}_{2}'\left(\kappai\right)}
    {\kappai}
   \left[
    14
    \dfrac{{Q}_{-3}\left(\kappae\right)}{\kappae^2}
 \right.
\right.
\\
\shoveright{
\left.
 \left.
    -
    4\dfrac{{Q}_{-3}'\left(\kappae\right)}{\kappae}
   \right]
\right.
}
\\
\left.
 +
 \dfrac{\etai}{\etae}
 {Q}_{2}''\left(\kappai\right)
 \left[
  5
  \dfrac{{Q}_{-3}\left(\kappae\right)}{\kappae^2}
  -
  \dfrac{{Q}_{-3}'\left(\kappae\right)}{\kappae}
 \right]
\right.
\\
\shoveleft{
\hspace*{3em}
\left.
 +
 \dfrac
  {{Q}_{2}'\left(\kappai\right)}
  {\kappai}
 \left[
  16
  \dfrac{{Q}_{-3}\left(\kappae\right)}{\kappae^2}
 \right.
\right.
}
\\
\shoveright{
\left.
 \left.
  -
  6\dfrac{{Q}_{-3}'\left(\kappae\right)}{\kappae}
  +
  {Q}_{-3}''\left(\kappae\right)
 \right]
\right\}
}
\\
\left(
  3
  \left\{
   \dfrac{\etai}{\etae}
   \dfrac
    {{Q}_{-3}'\left(\kappae\right)}
    {\kappae}
   \left[
    4\dfrac{{Q}_{2}'\left(\kappai\right)}{\kappai}
    +
    {Q}_{2}''\left(\kappai\right)
   \right]
 \right.
\right.
\\
\left.
 \left.
   +
   \dfrac
    {{Q}_{2}'\left(\kappai\right)}
    {\kappai}
   \left[
    6\dfrac{{Q}_{-3}'\left(\kappae\right)}{\kappae}
    -
    {Q}_{-3}''\left(\kappae\right)
   \right]
  \right\}
\right)^{-1}
,
\label{Ce21}
\end{multline}
\begin{multline}
\hat{B}_{\mathrm{i}2,1}
=
\dfrac{\mathrm{i}\omega\tau}{2}\hat{h}_{2,1,\omega}
\left[
   6\dfrac{\etai}{\etae}
   \dfrac
    {{Q}_{-3}'\left(\kappae\right)}
    {\kappae}
\right.
\\
\shoveright{
\left.
  +
  14
  \dfrac
   {{Q}_{-3}'\left(\kappae\right)}
   {\kappae}
  -
  5
  {Q}_{-3}''\left(\kappae\right)
\right]
}
\\
\left(
  3
  \left\{
   \dfrac{\etai}{\etae}
   \dfrac
    {{Q}_{-3}'\left(\kappae\right)}
    {\kappae}
   \left[
    4\dfrac{{Q}_{2}'\left(\kappai\right)}{\kappai}
    +
    {Q}_{2}''\left(\kappai\right)
   \right]
 \right.
\right.
\\
\left.
 \left.
   +
   \dfrac
    {{Q}_{2}'\left(\kappai\right)}
    {\kappai}
   \left[
    6\dfrac{{Q}_{-3}'\left(\kappae\right)}{\kappae}
    -
    {Q}_{-3}''\left(\kappae\right)
   \right]
  \right\}
\right)^{-1}
,
\label{Bi21}
\end{multline}
\begin{multline}
\hat{C}_{\mathrm{i}2,1}
=
-
\dfrac{\mathrm{i}\omega\tau}{2}\hat{h}_{2,1,\omega}
\left\{
   \dfrac{\etai}{\etae}
   \dfrac
    {{Q}_{-3}'\left(\kappae\right)}
    {\kappae}
   \left[
    6
    \dfrac{{Q}_{2}\left(\kappai\right)}{\kappai^2}
 \right.
\right.
\\*
\shoveright{
\left.
 \left.
    +
    4\dfrac{{Q}_{2}'\left(\kappai\right)}{\kappai}
    +
    {Q}_{2}''\left(\kappai\right)
   \right]
\right.
}
\\*
\left.
   +
   \dfrac
    {{Q}_{-3}'\left(\kappae\right)}
    {\kappae}
   \left[
    14
    \dfrac{{Q}_{2}\left(\kappai\right)}{\kappai^2}
    +
    6\dfrac{{Q}_{2}'\left(\kappai\right)}{\kappai}
   \right]
\right.
\\*
\shoveright{
\left.
 -
 {Q}_{-3}''\left(\kappae\right)
 \left[
  5\dfrac{{Q}_{2}\left(\kappai\right)}{\kappai^2}
  +
  \dfrac{{Q}_{2}'\left(\kappai\right)}{\kappai}
 \right]
\right\}
}
\\
\left(
  3
  \left\{
   \dfrac{\etai}{\etae}
   \dfrac
    {{Q}_{-3}'\left(\kappae\right)}
    {\kappae}
   \left[
    4\dfrac{{Q}_{2}'\left(\kappai\right)}{\kappai}
    +
    {Q}_{2}''\left(\kappai\right)
   \right]
 \right.
\right.
\\
\left.
 \left.
   +
   \dfrac
    {{Q}_{2}'\left(\kappai\right)}
    {\kappai}
   \left[
    6\dfrac{{Q}_{-3}'\left(\kappae\right)}{\kappae}
    -
    {Q}_{-3}''\left(\kappae\right)
   \right]
  \right\}
\right)^{-1}
,
\label{Ci21}
\end{multline}
\begin{equation}
\vec{M}_{\mathrm{e}1,1}
=
\dfrac{2}{5}\left(\mue-\mui\right)
\dfrac{13\mue+11\mui}{\left(\mui+2\mue\right)^2}
\Rd^2\vec{H}_\mathrm{a}\cdot\hat{h}_{2,1}
,
\end{equation}
\begin{equation}
\vec{M}_{\mathrm{i}1,1}
=
\dfrac{4}{5}\left(\mue-\mui\right)
\dfrac{7\mui+5\mue}{\left(\mui+2\mue\right)^2}
\Rd^2\vec{H}_\mathrm{a}\cdot\hat{h}_{2,1}
,
\end{equation}
\begin{equation}
\hat{M}_{\mathrm{e}3,1}
=
\dfrac{\mui-\mue}{\mui+2\mue}
\dfrac{5\mui-8\mue}{3\mui+4\mue}
\Rd^4
\left[
 \left(\vec{H}_\mathrm{a}\hat{h}_{2,1}\right)^\mathrm{S}
\right]^\mathrm{D}
,
\end{equation}
\begin{equation}
\hat{M}_{\mathrm{i}3,1}
=
-
4\dfrac{\mui-\mue}{\mui+2\mue}
\dfrac{\mui-\mue}{3\mui+4\mue}
\Rd^4
\left[
 \left(\vec{H}_\mathrm{a}\hat{h}_{2,1}\right)^\mathrm{S}
\right]^\mathrm{D}
,
\end{equation}
where
\vspace*{-2ex}
\begin{equation}
\tau
=
\dfrac{4\etae\Rd}{\sigmas}
,
\end{equation}
\begin{equation}
\hat{h}_{2,1}
=
\hat{h}_{2,1}(t)
=
\dfrac{1}{2\pi}
\int\limits_{-\infty}^\infty
 \en^{-\mathrm{i}\omega{t}}
 \hat{h}_{2,1,\omega}
\mathrm{d}\omega
,
\end{equation}
\begin{equation}
\left[
 \left(\vec{H}_\mathrm{a}\hat{h}_{2,1}\right)^\mathrm{S}
\right]^\mathrm{D}
=
\left(\vec{H}_\mathrm{a}\hat{h}_{2,1}\right)^\mathrm{S}
 -
\dfrac{3}{5}
 \left(\vec{H}_\mathrm{a}\cdot\hat{h}_{2,1}\hat{I}\right)^\mathrm{S}
,
\end{equation}
and $\hat{h}_{2,1,\omega}$ is the solution of the following equation
\begin{multline}
\dfrac{1}{2\pi}
\int\limits_{-\infty}^\infty
 \en^{-\mathrm{i}\omega{t}}
 \left[
  1
  -
  \dfrac{\mathrm{i}\omega\tau}{48}
  {F}(\kappae,\kappai)
 \right]
 \hat{h}_{2,1,\omega}
\mathrm{d}\omega
\\
=
\dfrac
 {\left(\mui-\mue\right)^2\Rd}
 {\left(2\mue+\mui\right)^2{H}_\mathrm{am}^2}
\left(\vec{H}_\mathrm{a}^2-\dfrac{1}{3}\hat{I}\Ha^2\right)
,
\label{eq:h21}
\end{multline}
\begin{multline}
{F}(\kappae,\kappai)
\\
=
\dfrac
 {P_\mathrm{Ns}(\kappae,\kappai^2)\sinh\kappai
  +
  P_\mathrm{Nc}(\kappae,\kappai^2)\kappai\cosh\kappai}
 {P_\mathrm{Ds}(\kappae,\kappai^2)\sinh\kappai
  +
  P_\mathrm{Dc}n(\kappae,\kappai^2)\kappai\cosh\kappai}
,
\end{multline}
\begin{multline}
P_\mathrm{Ds}(\kappae,\kappai^2)
=
(\kappai^4+15\kappai^2+30)(\kappae+1)\dfrac{\etai}{\etae}
\\
-
3(2\kappai^2+5)(\kappae^2+5\kappae+5)
,
\end{multline}
\begin{multline}
P_\mathrm{Dc}(\kappae,\kappai^2)
=
-
5(\kappai^2+6)(\kappae+1)\dfrac{\etai}{\etae}
\\
+
(\kappai^2+15)(\kappae^2+5\kappae+5)
,
\end{multline}
\begin{multline}
P_\mathrm{Ns}(\kappae,\kappai^2)
=
3(\kappai^6+25\kappai^4+252\kappai^2+480)
(\kappae+1)\dfrac{\etai^2}{\etae^2}
\\
+
(
 2\kappae^3\kappai^4
 +
 9\kappae^2\kappai^4
 +
 30\kappae^3\kappai^2
 -
 45\kappae\kappai^4
\\
 +
 540\kappae^2\kappai^2
 -
 45\kappai^4
 +
 60\kappae^3
 +
 198\kappae\kappai^2
\\
 +
 1260\kappae^2
 +
 198\kappai^2
 +
 720\kappae
 +
 720
)
\dfrac{\etai}{\etae}
\\
-
6(2\kappai^2+5)
(\kappae^4+5\kappae^3+45\kappae^2+72\kappae+72)
,
\end{multline}
\begin{multline}
P_\mathrm{Nc}(\kappae,\kappai^2)
=
-
3(5\kappai^4+92\kappai^2+480)(\kappae+1)\dfrac{\etai^2}{\etae^2}
\\
-
(
 -
 3\kappae^2\kappai^4
 +
 10\kappae^3\kappai^2
 -
 15\kappae\kappai^4
 +
 120\kappae^2\kappai^2
\\
 -
 15\kappai^4
 +
 60\kappae^3
 -
 42\kappae\kappai^2
 +
 1260\kappae^2
\\
 -
 42\kappai^2
 +
 720\kappae
 +
 720
)
\dfrac{\etai}{\etae}
\\
+
2(\kappai^2+15)
(\kappae^4+5\kappae^3+45\kappae^2+72\kappae+72)
.
\end{multline}
Here, $\mathrm{i}$ is the imaginary unit, $\sinh$ and $\cosh$ are the hyperbolic sine and cosine.

The obtained relations for the terms of the asymptotic expansions allow one also to find the expressions for the corresponding unknown scalar, vector, and tensor functions for the free proper oscillations of the drop. They have, in general, infinite number of modes determined by the initial disturbances of the spherical shape and decay due to the viscosity. These expressions are too cumbersome and are not written down here.

\begin{figure}
\centerline{\includegraphics{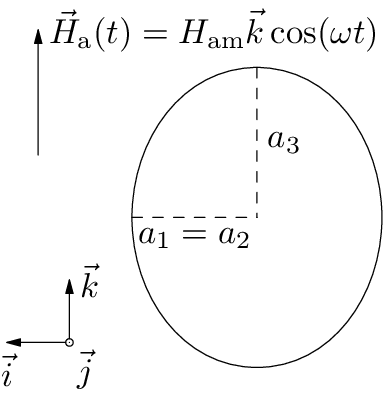}}
\caption{Drop in an applied uniform harmonically oscillating magnetic field.}
\label{f1}
\end{figure}
With accuracy up to the terms of the first order, in an oscillating applied magnetic field with the intensity 
\begin{equation}
\vec{H}_\mathrm{a}
=
H_\mathrm{am}\cos\left(\omega{t}\right)\vec{k}
, 
\end{equation}
the drop is a prolate spheroid with the axis directed along $\vec{k}$ (see Fig.~\ref{f1}) and with the semiaxes 
\begin{multline}
a_1=a_2
\\
=
\Rd
\left\{
 1
 -
 \dfrac{\alpha}{6}
 \left(\dfrac{\mui-\mue}{\mui+2\mue}\right)^2
 \left[
  1
  +
  \left|\zeta\right|\cos\left(2\omega{t}-2\phi\right)
 \right]
\right\}
,
\end{multline}
\begin{equation}
a_3
=
\Rd
\left\{
 1
 +
 \dfrac{\alpha}{3}
 \left(\dfrac{\mui-\mue}{\mui+2\mue}\right)^2
 \left[
  1
  +
  \left|\zeta\right|\cos\left(2\omega{t}-2\phi\right)
 \right]
\right\}
,
\end{equation}
where
\begin{equation}
\zeta
=
\left[
 1
 -
 \dfrac{i\omega\tau}{96}{F}(\kappae,\kappai)
\right]^{-1}
,
\end{equation}
\begin{equation}
\phi
=
\dfrac{1}{2}
\arctan
 \left(
  -
  \dfrac{\mathop\mathrm{Im}\zeta}{\mathop\mathrm{Re}\zeta}
 \right)
.
\end{equation}
Here, $\mathrm{Re}$ and $\mathrm{Im}$ denote the real and imaginary parts of a complex quantity, $\arctan$ is the arc tangent.  Thus, the drop performs deformational oscillations with the angular frequency $2\omega$ and the phase lag $2\phi$.

\begin{figure}
\centerline{\includegraphics{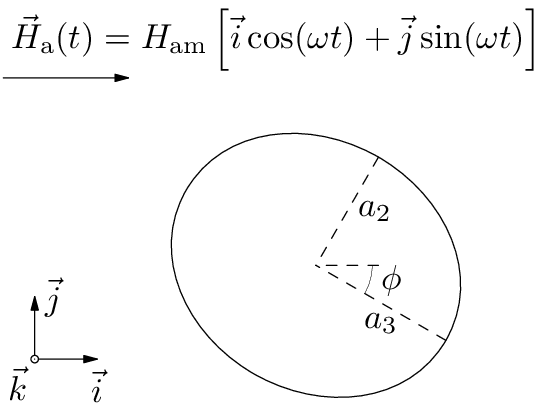}}
\caption{Drop in an applied uniform rotating magnetic field.}
\label{f2}
\end{figure}
\begin{figure*}
\includegraphics{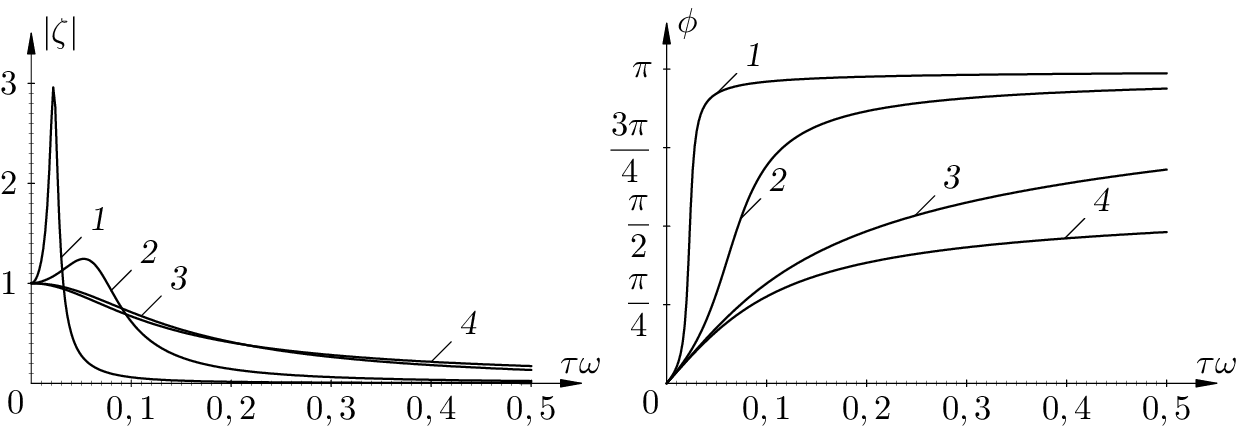}
\caption{The dependencies of $\left|\zeta\right|$ and $\phi$ upon the angular frequency $\omega$ for $\etai=20~\mathrm{mPa}~\mathrm{s}$, $\rhoi=1.4~\mathrm{kg}/\mathrm{m}$, $\etae=0.89~\mathrm{mPa}~\mathrm{s}$, $\rhoe=1~\mathrm{kg}/\mathrm{m}$, $\sigmas=20~\mathrm{mN}/\mathrm{m}$ (1:~$\Rd=1~\mathrm{mm}$, 2:~$\Rd=100$~\textmu$\mathrm{m}$, 3:~$\Rd=10$~\textmu$\mathrm{m}$, 4:~$\Rd=1$~\textmu$\mathrm{m}$).}
\label{fig}
\end{figure*}
With accuracy up to the terms of the first order, in a rotating applied magnetic field with the intensity 
\begin{equation}
\vec{H}_\mathrm{a}
=
H_\mathrm{am}
\left[
 \cos\left(\omega{t}\right)\vec{i}
 +
 \sin\left(\omega{t}\right)\vec{j}
\right]
, 
\end{equation}
the drop takes the shape of a tri-axial ellipsoid (see Fig.~\ref{f2}) with the semiaxes
\begin{align}
a_1
&=
\Rd
\left[
 1
 -
 \dfrac{\alpha}{3}
 \left(\dfrac{\mui-\mue}{\mui+2\mue}\right)^2
\right]
,
\\
a_2
&=
\Rd
\left[
 1
 +
 \dfrac{\alpha}{6}
 \left(\dfrac{\mui-\mue}{\mui+2\mue}\right)^2
 \left(
  1
  -
  3\left|\zeta\right|
 \right)
\right]
,
\\
a_3
&=
\Rd
\left[
 1
 +
 \dfrac{\alpha}{6}
 \left(\dfrac{\mui-\mue}{\mui+2\mue}\right)^2
 \left(
  1
  +
  3\left|\zeta\right|
 \right)
\right]
.
\end{align}
The ellipsoid rotates around its minor axis, directed along $\vec{k}=\vec{i}\times\vec{j}$, with the angular speed $\omega$ so that its major axis lags from $\vec{H}_\mathrm{a}$ by the angle $\phi$. Here, $\vec{i}$, $\vec{j}$, and $\vec{k}$ form a right triple of orthonormal vectors.

The dependencies of $\left|\zeta\right|$ and $\phi$ upon $\omega$ for a drop of magnetic liquid on the basis of magnetite in kerosene suspended in water at various values of the radius of the drop, $\Rd$, are presented in Fig.~\ref{fig}.

\subsection{Limit cases.}

\begin{figure*}
\includegraphics{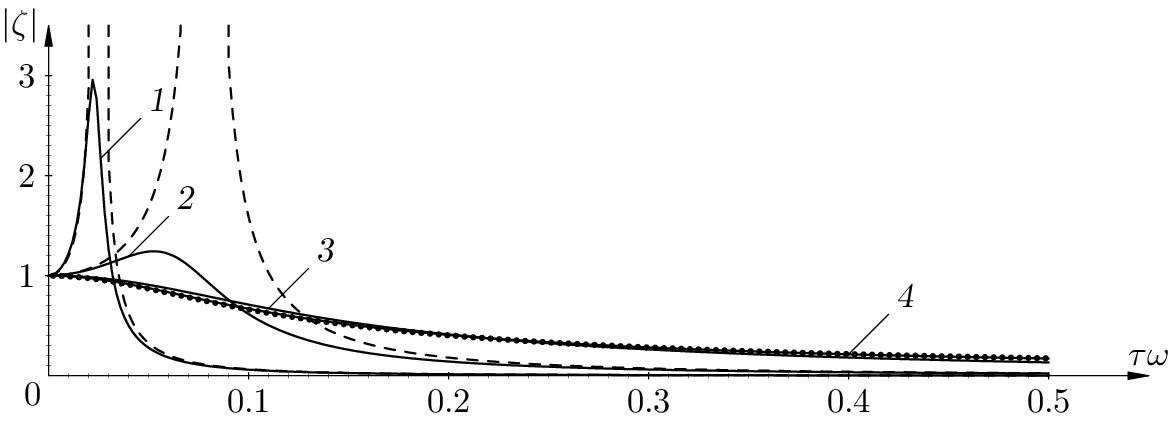}
\caption{The dependencies of $\left|\zeta\right|$ upon the angular frequency $\omega$ for $\etai=20~\mathrm{mPa}~\mathrm{s}$, $\rhoi=1.4~\mathrm{kg}/\mathrm{m}$, $\etae=0.89~\mathrm{mPa}~\mathrm{s}$, $\rhoe=1~\mathrm{kg}/\mathrm{m}$, $\sigmas=20~\mathrm{mN}/\mathrm{m}$ (1:~$\Rd=1~\mathrm{mm}$, 2:~$\Rd=100$~\textmu$\mathrm{m}$, 3:~$\Rd=10$~\textmu$\mathrm{m}$, 4:~$\Rd=1$~\textmu$\mathrm{m}$) as compared with those for the limit cases: the dotted line when the inertia is neglected and the dashed lines when the viscosity is neglected.}
\label{f3}
\end{figure*}
For the limit case $\rhoe\sigmas\Rd/(2\etae^2)\to0$, 
\begin{multline}
\left|\zeta\right|
\to
\dfrac{1}{\sqrt{1+4\tau_{2,1}^2\omega^2}}
,\quad
\phi
\to
\dfrac{1}{2}\arctan\left(2\tau_{2,1}\omega\right)
\\
\text{ as } 
\dfrac{\rhoe\sigmas{a}}{2\etae^2}\to0
,
\end{multline}
\begin{equation}
\tau_{2,1}
=
\dfrac
 {\left(16\etae+19\etai\right)\left(3\etae+2\etai\right)\Rd}
 {40\left(\etae+\etai\right)\sigmas}
.
\end{equation}
This case corresponds to the case when the inertia of the liquids can be neglected, considered in \cite{EPJ,MHD18,BRAS-Ph}.

For the limit case $\rhoe\sigmas\Rd/(2\etae^2)\to\infty$,
\begin{multline}
\left|\zeta\right|
\to
\dfrac{\omega^2_{2,1}}{\left|\omega^2_{2,1}-4\omega^2\right|}
,\quad
\phi
\to
\begin{cases}
0,\quad 2\omega<\omega_{2,1}
\\
\pi, \quad 2\omega>\omega_{2,1}
\end{cases}
\\
\text{ as } 
\dfrac{\rhoe\sigmas{a}}{2\etae^2}\to\infty
,
\end{multline}
\begin{equation}
\omega_{2,1}
=
\sqrt{\dfrac{24\rhoe}{2\rhoe+3\rhoi}
 \dfrac{\sigmas}{\rhoe\Rd^3}}
.
\end{equation}
This case corresponds to the case when the viscosity of the liquids can be neglected, considered in \cite{FD21}. The limit resonance circular frequency $\omega_{2,1}$ coincides with the frequency of the proper free deformational oscillations (for the mode with $n=2$) found by Rayleigh (see \cite{Lamb} art.~275).

In Fig.~\ref{f3}, the dependencies of $\left|\zeta\right|$ upon the angular frequency $\omega$ are presented together with those  for the limit cases. As can be seen from the figure, the inertia may be neglected for small oscillations of a drop of a typical magnetic liquid suspended in water if the radius of the drop does not exceed 10 microns. And, for sufficiently large-scale drops, the formulas in which the influence of the viscosity is neglected may be used only if the frequency of the oscillations of the applied field is sufficiently far from the resonance frequency.

\section{Conclusion}

The behavior of a drop of a viscous magnetic liquid suspended in another viscous magnetic liquid immiscible with the former in an applied non-stationary uniform magnetic field is theoretically investigated within the approximation of low Reynolds numbers with taking into account that the flow of the liquids is unsteady. The general solution of the Navier--Stokes equation in the low Reynolds number approximation is found for unsteady flows of an incompressible liquid. With the use of this general solution in the first order approximation over the small parameter smallness of which provides the smallness of the deformation of the drop, the equations are obtained that describe the variation of the form of the drop in an arbitrarily varying uniform magnetic field. The solutions of these equations are found for forced steady-state oscillations of the drop in harmonically oscillating and rotating applied magnetic fields.

As the solution of the given problem shows, rather cumbersome expressions even for the corrections of the first order appear in taking into account the unsteady term. So more careful study of the general solution with the goal to find more simple asymptotic expressions for the case when the viscosity tends to zero acquires actuality. Those asymptotic expressions should allow describing the process of the oscillation decay and finiteness of the oscillation amplitude for the resonance.

\begin{acknowledgments}
The partial support by RFBR grant 19-01-00056 is acknowledged.
\end{acknowledgments}

\end{document}